\documentclass[twocolumn,aps,prl,showpacs]{revtex4}
\usepackage{amsmath}
\usepackage{amssymb}
\usepackage{color}
\usepackage[dvips]{graphicx}

\begin{document}

\title{Broadening the bandwidth of entangled photons: a step towards the generation of extremely short biphotons}

\author{M. Hendrych$^1$, Xiaojuan Shi$^{1,2}$, A. Valencia$^1$, and Juan P. Torres$^{1,2}$}

\affiliation{$^1$ICFO-Institut de Ciencies Fotoniques,
Mediterranean Technlogy Park, Castelldefels, 08860 Barcelona,
Spain}
\affiliation{$^2$Department of Signal Theory and
Communications, Campus Nord D3, Universitat Politecnica, 080340
Barcelona, Spain} \email{martin.hendrych@icfo.es}

\date{\today}
\begin{abstract}
We demonstrate a technique that allows to fully control the
bandwidth of entangled photons independently of the frequency band
of interest and of the nonlinear crystal. We show that this
technique allows to generate nearly transform-limited biphotons
with almost one octave of bandwidth (hundreds of THz) which
corresponds to correlation times of just a few femtoseconds. The
presented method becomes an enabling tool for attosecond
entangled-photons quantum optics. The technique can also be used
to generate paired photons with a very high degree of
entanglement.
\end{abstract}


\maketitle

The development of methods for the generation of entangled photon
pairs (biphotons) with a specific bandwidth has been of great
interest in recent years. Narrow bandwidth of frequency
correlations is important for the design of efficient atom-photon
interfaces \cite{atomphoton} in quantum networks~\cite{DLCZ}, for
long distance quantum communications~\cite{ZeilKarlsson}, or to
enable direct measurements of temporal correlations with current
photodetectors~\cite{harris1,ou1}.

On the other hand, some applications such as quantum optical
coherence tomography~\cite{nasr1} and nonlinear
microscopy~\cite{Silberberg} require wide bandwidths. Wide
bandwidths are a requisite for the generation of biphotons with
very short correlation times~\cite{harris2} and when high fluxes
of biphotons are desired~\cite{Loudon}. A bandwidth of hundreds of
THz can generate biphotons with a few femtoseconds of correlation
time. These short temporal biphotons are of particular interest in
the fields of quantum metrology~\cite{QMetrology} and for some
protocols for timing and positioning
measurements~\cite{valencia1}. The narrow temporal correlation
embedded in the biphoton can be transmitted over large distances
thanks to the strong correlations of the entangled photons that
allow to remotely compensate for chromatic
dispersion~\cite{masha}.

Spontaneous parametric down conversion (SPDC) is the most
convenient source for the generation of entangled photon pairs.
When an intense laser beam illuminates a nonlinear material, with
a certain small probability a pump photon might split into two
lower energy photons. In analogy with a classical pulse, the
temporal correlation width of the biphoton is determined by its
spectral bandwidth and spectral phase~\cite{rubin1}. The bandwidth
of the SPDC pairs is determined by the type of phase-matching
(type-I, type-II), the length and the dispersive properties of the
nonlinear material, the geometry of the SPDC configuration
(collinear or non-collinear) and by the spectral and spatial
characteristics of the pump beam. By changing any of these
parameters it is possible to modify the bandwidth and therefore
the temporal correlations of the SPDC photons. For example, by
choosing a particular material (PPLN) and pumping it at a specific
wavelength ($\lambda =1885~{\rm nm}$), an ultra broad bandwidth of
$\sim 1080~{\rm nm}$ has been recently reported~\cite{kevin1}.
However, from this experiment no conclusion can be drawn about the
biphoton's temporal length because the spectral correlations were
not measured. In contrast, the group of Silberberg achieved to
produce a bandwidth of 100~nm that yielded a biphoton with 23 fs
of correlation time \cite{Silberberg}.

An alternative approach to modify the SPDC bandwidth is to use
chirped quasi-phase matched crystals~\cite{nasr2}. A frequency
correlation bandwidth of $\sim 300~{\rm nm}$ at a central
wavelength of $812~{\rm nm}$ has been achieved. Nevertheless, when
this technique is used to obtain narrow temporal biphotons,
 compensation of the spectral phase is
required as the biphotons are not transform limited.

In this Letter, we demonstrate experimentally a technique to
increase the SPDC bandwidth in order to generate ultrashort near
transform-limited biphotons. It employs angular dispersion
(pulse-front tilt) to modify the effective group velocity and
group velocity dispersion of the interacting waves. Differently
from other approaches, the method is not based on the choice nor
on the engineering of the nonlinear material and it works in any
frequency band and in any nonlinear crystal of interest. That
offers two advantages: First, the wavelength can be chosen in the
region where single-photon detectors exhibit high detection
efficiency; second, advantage can be taken of materials with high
nonlinear coefficient that naturally do not provide the desired
bandwidth.

Angular dispersion is a powerful enabling tool in many different
areas of optics. It can be used to introduce negative group
velocity dispersion in a beam propagating in free space
\cite{gordon1}, it is a key element in many techniques for pulse
compression, it enables the generation of femtosecond second
harmonic waves \cite{volosov1,dubietis1}, and it has made possible
the observation of temporal solitons \cite{DiTrapani}, where the
natural dispersion of the material would have made such
observation impossible. In quantum optics, angular dispersion
enables to tailor the frequency correlations of entangled photons
and allows to generate frequency-uncorrelated and
frequency-correlated pairs of photons \cite{martin1}, quantum
states that are not easily produced in most experimental
configurations currently used \cite{wong1,mosley1}.


Let us consider collinear SPDC where the downconverted photons
copropagate along the direction of the pump beam. The state of the
downconverted photons at the output of the medium may be written
as
\begin{equation}\label{spdcstate}
\left|\psi \right\rangle = \int d{\bf q}_s d{\bf q}_i d\Omega_{s}
d\Omega_{i}\Phi \left( \Omega_{s},\Omega_{i},{\bf q}_s,{\bf q}_i
\right)\left|\Omega_{s}\right\rangle\left|\Omega_{i}\right\rangle,
\end{equation}
where $\Omega_{j}$ ($j=s,i$) denote the signal and idler photons'
angular frequency detunings from the central angular frequency
$\omega_{j}^{0}$, i.e, $\omega_{j}= \omega_{j}^{0}+\Omega_{j}$,
and ${\bf q}_j$ are the corresponding transverse wavenumber
vectors. The spectral and spatial properties of the down-converted
photons are described by the joint spectrum~\cite{rubin1}
\begin{eqnarray}\label{jointspectrumtheory}
\Phi \left( \Omega_{s},\Omega_{i},{\bf q}_s,{\bf q}_i \right) &
\propto & E_{\omega} \left( \Omega_s+\Omega_i \right) E_{q}
\left( {\bf q}_s+{\bf q}_i \right) \nonumber \\
& \times & \mathrm{sinc} \left( \frac{\Delta k L}{2} \right) \exp
\left\{i\frac{s_k L}{2} \right\},
\end{eqnarray}
where $E_{\omega}$ and $E_{q}$ are the spectral and transverse
wavenumber distributions of the pump beam at the input face of the
nonlinear crystal of length $L$, respectively. $\Delta k=
k_{p}-k_{s}-k_{i}$ is the phase-mismatch along the longitudinal
direction with $k_{j}=[\left( \omega_{j} n_{j} \right)^2/c^2-|{\bf
q}_{j}|^{2}]^{1/2}$, and $s_k=k_p+k_s+k_i$. $n_{j}$ is the index
of refraction and $c$ the speed of light.

\begin{figure}
\centering\includegraphics[scale=0.85,width=0.9\columnwidth]{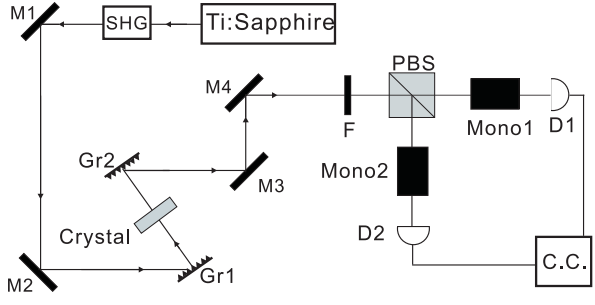}
\caption{Experimental setup. SHG: second harmonic generation; M:
mirrors; Gr: gratings; F: long-pass filter; PBS: polarizing
beamsplitter; Mono: monochromators; D: detectors; C.C.
Coincidences electronics.} \label{setup}
\end{figure}

To illustrate the principle behind the proposed method, let us
consider the scheme shown in Fig.~\ref{setup}. In contrast to
typical collinear SPDC configurations, the nonlinear crystal is
placed between two optical elements that introduce angular
dispersion, e.g., a pair of prisms or diffraction gratings.
Angular dispersion $\epsilon$ causes the front of the pulse to be
tilted by an angle $\xi$ given by $\tan \xi=-\lambda_{0}
\epsilon$, where $\lambda_{0}$ is the central wavelength of the
pulse and angular dispersion $\epsilon=m/\left( d \cos \beta_0
\right)$, where $m$ is the grating's diffraction order, $d$ is the
groove spacing and $\beta_0$ is the output diffraction angle.

The introduction of angular dispersion modifies the spatial
distribution of the pump $E_{q}$ and the phase-mismatch $\Delta k$
of the joint spectrum of Eq.~\ref{jointspectrumtheory}. If the
gratings introduce opposite angular dispersion in such a way that
$\tan\xi_{s}=-\tan\xi_{p}/\alpha_{p}$ and
$\alpha_{p}\alpha_{s}=1$, it is possible to modify $\Delta k$
only, and to effectively modify the group velocity and the group
velocity dispersion of the pump, signal and idler photons. Here
$\alpha_{j}\equiv\cos\theta_{j}/\cos\beta_{j}$ with $\theta_{j}$
and $\beta_{j}$ being the incidence and diffraction angles of the
pump beam at grating $Gr1$, respectively.

The effects of angular dispersion on the SPDC spectrum can be
better understood by expanding $\Delta k$ in a Taylor series about
the central frequencies $\omega_{j}^{0}$. For a
quasi-continuous-wave pump, the frequency detunings of the
downconverted photons must be anticorrelated in order to satisfy
energy conservation, i.e. $\Omega_{s}=-\Omega_{i}$, and $\Delta k$
becomes
\begin{equation}\label{PM2order}
\Delta k\approx \left( N_{s}^{\prime}-N_{i}^{\prime}
\right)\Omega_{s} -\frac{1}{2} \left(
g_{s}^{\prime}+g_{i}^{\prime} \right)\Omega_{s}^{2}+ \dots ,
\end{equation}
where $N_{j}^{\prime}=N_j+ \tan {\xi_{p}} \tan {\rho_{j}} / c$ and
$g^{\prime}_{j}=g_j-\left[\tan \xi_{p}/c\right]^{2}/k_{j}^{0}$
play the role of effective inverse group velocity and effective
group velocity dispersion, respectively. The inverse group
velocity $N_j = \left( d k_{j}/d \omega_{j} \right)_{\omega_j^0}$
and group velocity dispersion $g_j = \left( d^{2}k_{j} /d
\omega^{2}_{j} \right)_{\omega_{j}^{0}}$ are modified in the
presence of the pulse-front tilt $\xi_{p}$ and Poynting-vector
walk-off $\rho_{j}$. The new effective values of group velocity
and group velocity dispersion depend on the angular dispersion
experienced by the pump beam and allow us to control the SPDC
bandwidth.

For example, let us consider type-II SPDC where the polarizations
of the downconverted photons are mutually orthogonal. If tilt
$\xi_{p}$ is chosen such that the effective group velocities are
equal, $N_{s}^{\prime}=N_{i}^{\prime}$, the lowest non-zero term
in phase mismatch $\Delta k$ is of the second order which results
in an increase of the bandwidth. We obtain a type-II process where
the dependence of the bandwidth on the length of the nonlinear
crystal goes as $1/\sqrt{L}$, instead of the typical $1/L$
\cite{rubin1}. The value of the pulse tilt that maximizes the
bandwidth is
\begin{equation} \label{maxtiltangleII}
\xi^{max}_{II} = \tan ^{-1} \left \{\frac{c
(N_{i}-N_{s})}{\tan \rho_{s}-\tan \rho_{i}}\right\}.
\end{equation}

On the other hand, in a type-I process the polarizations and the
group velocities of the signal and idler photons are equal. The
bandwidth is increased if the tilt is chosen such that
$g_{s}^{\prime}=g_{i}^{\prime}=0$. In such a case, the first
non-zero term in $\Delta k$ is of the $4^{\rm th}$ order and the
dependence of the bandwidth on the length of the crystal goes as
$1/L^{1/4}$. The tilt that maximizes the bandwidth is
\begin{equation} \label{maxtiltangleI}
\xi^{max}_{I} = \tan ^{-1} \left( c^{2} g_{s} k_{s}^{0}
\right)^{1/2},
\end{equation}
where $k_{j}^{0}=k_{j}\left( \omega_{j}^{0},{\bf q}_s={\bf q}_i=0
\right)$.

\begin{figure}
\centering\includegraphics[scale=1,width=1\columnwidth]{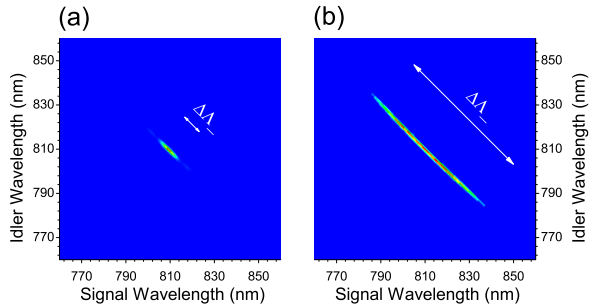}
\caption{Measured joint spectral density: (a) tilt $\xi_{p} = 0
^{\circ}$, (b) tilt $\xi_{p} = 38^\circ$. The joint spectrum
broadens sevenfold as expected.} \label{jointspec}
\end{figure}


To demonstrate the feasibility of the proposed technique, an
experiment was performed (see Fig.~\ref{setup}). A 2~mm thick,
type-II BBO crystal flanked with two gratings $Gr1$ and $Gr2$ was
pumped by the second harmonic (Radiantis Blue Stream, 40\%
efficiency in the picosecond regime) of a picosecond Ti:sapphire
laser (Coherent Mira 900-P) tuned at 810 nm. The downconverted
photons were separated by a polarizing beamsplitter and fed into
monochromators $Mono1$ and $Mono2$ (Jobin Yvon MicroHR).
Single-photon detectors (Perkin-Elmer SPCM-AQR-14-FC) measured
single counts and coincidence electronics counted coincidence
counts within a 3~ns window. The joint spectrum was obtained by
measuring the number of coincidences while the two monochromators
scanned the plane from 760 nm to 860~nm.

Figure~\ref{jointspec} shows the experimental results.
Fig.~\ref{jointspec}(a) depicts the joint spectral density for
tilt $\xi_{p}=0^{\circ}$ when the gratings were removed. The
typical frequency anti-correlation of the SPDC photons can be
observed. Fig.~\ref{jointspec}(b) corresponds to the situation
with tilt $\xi^{max}_{II}=38^\circ$ that maximizes the bandwidth.
This was accomplished by taking grating $Gr1$ with a groove
spacing $d=1/1200$ mm and diffraction angle $\beta_{p}=52^\circ$.
In order to satisfy $\tan\xi_{s}=-\tan\xi_{p}/\alpha_{p}$ and
$\alpha_{p}\alpha_{s}=1$, grating $Gr2$ had $d=1/600$ mm and
$\beta_{s}=18^\circ$.

\begin{figure}
\centering\includegraphics[scale=1,width=1\columnwidth]{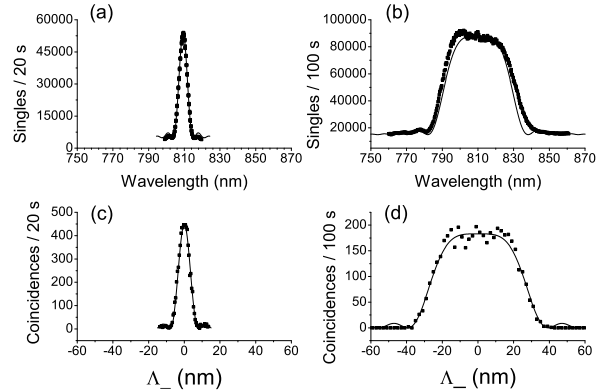}
\caption{(a) Signal single counts for $\xi_{p}=0^\circ$; $\Delta
\lambda_{s}=5.2~{\rm nm}$. (b) Signal single counts for
$\xi_{p}=38^\circ$; $\Delta \lambda_{s}=41~{\rm nm}$. (c)
Coincidences along the antidiagonal for $\xi_{p}=0^\circ$; $\Delta
\Lambda_{-}=7.5~{\rm nm}$. (d) Coincidences along the antidiagonal
for $\xi_{p}=38^\circ$; $\Delta \Lambda_{-}=52~{\rm nm}$. Solid
lines represent the theoretical prediction, squares are the
experimental data.} \label{spectra}
\end{figure}

A more quantitative comparison between the cases with and without
pulse-front tilt is plotted in Fig.~\ref{spectra}. The graphs in
the upper row correspond to the spectra of signal single counts
measured after one of the monochromators: (a) without tilt and (b)
with tilt $\xi_{p}=38^\circ$. The spectra of idler photons are not
shown for being alike. The squares are experimental data and the
solid lines are the theoretical prediction.

Figures~\ref{spectra}(c) and (d) depict the profile of the number
of coincidence counts along the antidiagonal (straight line at
$-45^\circ$) of Fig.~\ref{jointspec} without tilt and with tilt,
resp. The variable
$\Lambda_{-}=(\Lambda_{s}-\Lambda_{i})/\sqrt{2}$, where
$\Lambda_{j}$ is the detuning from the central wavelength, is
associated with this antidiagonal. Without tilt, the FWHM
bandwidth was measured to be $\Delta\Lambda_{-}\sim 7.5~{\rm nm}$
(Fig.~\ref{spectra}(c)). Applying a tilt $\xi_{p}=38^\circ$, the
bandwidth broadened to $\Delta\Lambda_{-}\sim 52~{\rm nm}$
(Fig.~\ref{spectra}(d)). The sevenfold increase of the bandwidth
is achieved without any modification of the nonlinear crystal or
the working wavelength, only by introducing angular dispersion
into the pump beam and the downconverted photons. It is this
independence of the material properties and of the working
wavelength which makes the method proposed here so promising for
controlling the bandwidth of the SPDC photons.

\begin{figure}
\centering\includegraphics[scale=1,width=1\columnwidth]{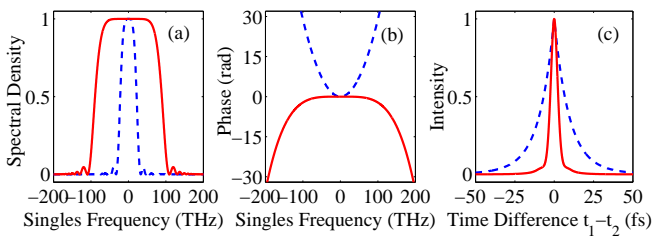}
\caption{ (a) Spectral density $|\Phi|^2$, (b) phase $s_k L/2$ and
(c) correlation time $|\Psi(t_1-t_2)|^2$ of the biphoton for
type-I BBO crystal without tilt (dashed line) and with tilt (solid
line). The shrinking of the temporal correlation is clearly
observed.} \label{biphoton}
\end{figure}

The application of this method for the generation of entangled
photons with very narrow temporal correlations can be seen in
Fig~\ref{biphoton}. In general, temporal biphoton $\Psi(t_1,t_2)$
($t_j$ is the clicking time of {\emph j}-th detector) is given by
the amplitude and phase of the joint spectrum $\Phi \left(
\Omega_{s},\Omega_{i} \right)$. Fig.~\ref{biphoton} shows the
spectral density~(a), spectral phase~(b) and temporal shape~(c) of
a biphoton for the case of 2-mm thick Type-I BBO SPDC at 810~nm.
Type-I phase matching was chosen because of its naturally broader
spectrum. In the case without tilt, the spectral phase follows a
quadratic dependence. In the case with tilt in the region where
the spectral density varies, the spectral phase is almost constant
due to its fourth-order dependence on the frequency. It is this
fact that makes possible to generate nearly transform-limited
biphotons. The mentioned increase of bandwidth translates into
ultrashort biphotons with a temporal correlation of a few
femtoseconds. This contrasts with other methods where the increase
of the bandwidth is not directly accompanied by a decrease of the
correlation time~\cite{WhiteLight,nasr2}.

Figure~\ref{biphoton}(c) was obtained by performing a Fourier
transform of Eq.~\ref{jointspectrumtheory}, both numerically and
analytically. The dashed and solid lines correspond to the case
${\xi_{p}=0^\circ}$ and ${\xi_{p}=\xi_{I}^{max}=16.2^\circ}$,
resp. The FWHM bandwidth of the spectral density obtained without
tilt is 96~nm ($2 \pi \, 44$~THz) that corresponds to a
correlation time of 19~fs (rms width) [13.4~fs FWHM]. This should
be compared with the case when the optimum angular dispersion is
applied that gives a bandwidth of 465~nm ($2\pi \, 197$~THz) and a
correlation time of 6.4~fs (rms width) [4.6~fs FWHM]. It should be
noted that working at a shorter wavelength, it is possible to
further shorten the biphoton, because the correlation time scales
with $\lambda^2/(c\Delta\lambda)$.

Pump beams with pulse-front tilt can also be employed to obtain a
very high degree of entanglement. Up to now low values of the
entropy of entanglement ($E \sim 1-2$ ebits) of paired photons in
the frequency continuum have been reported \cite{law1}. A higher
degree of entanglement can be achieved using broadband SPDC
photons. For entangled photons of the form $\Phi \left(
\Omega_s,\Omega_i \right) \sim \exp \left\{-\left(
\Omega_s+\Omega_i \right)^2/B_p^2 \right\}\exp \left\{-\left(
\Omega_s-\Omega_i \right)^2/B_c^2 \right\}$, the entropy of
entanglement \cite{parker1} depends on the ratio between the
bandwidth of the pump beam $B_p$ (typically $2\pi \, 5$ MHz) and
the bandwidth of the biphoton $B_c$. For a bandwidth of $\Delta
\lambda_s =31$ nm ($B_c \sim 2\pi \, 16.4$ THz) at $1064$ nm
\cite{silberberg1}, one has values of $B_c/B_p \sim 3.3 \times
10^6$ and $E \sim 21$ ebits. The method presented here allows to
reach values of $\Delta \lambda_s > 500$ nm ($B_c
> 2\pi \, 420$ THz), therefore allowing typical ratios greater than
$8.4 \times 10^7$ and $E>26$ ebits.

In conclusion, a scheme to broaden the bandwidth of the SPDC
photons was experimentally demonstrated. The method is based on
the introduction of angular dispersion (pulse-front tilt) into the
pump and downconverted photons which allowed us to demonstrate a
sevenfold increase of the original bandwidth without changing the
nonlinear crystal or the frequency band. The potentiality of the
proposed method to generate ultrashort nearly transformed-limited
biphotons, and with a high degree of entanglement, has also been
discussed. The described method can allow biphotons to enter the
realm of {\em attosecond quantum optics}, since it allows the
generation of entangled photons with correlation times below one
femtosecond.

Acknowledgements: We wish to thank Ladislav Mi\v{s}ta for the
analytical solution of the Fourier transform. This work was
supported by the European Commission [Qubit Applications, contract
015848] and by the Government of Spain [Consolider Ingenio 2010
(Quantum Optical Information Technology) CSD2006-00019 and
FIS2007-60179]. M. H. acknowledges support from a Beatriu de Pinos
fellowship.

\end{document}